\documentclass{article}
\usepackage{amssymb,hyperref}
\usepackage[english]{babel}
\usepackage{amsmath, graphics}
\usepackage{geometry}
\newcommand{\mathsym}[1]{{}}
\newcommand{\unicode}[1]{{}}

\def\noi{\noindent}
\def\nq{\hspace*{-1em}}
\def\nqq{\hspace*{-2em}}
\def\nn{\nonumber\\ {}}

\def\barr{\left(\begin{array}}
\def\earr{\end{array}\right)}
\def\beq#1{\begin{equation}\label{#1}}
\def\eeq{\end{equation}}
\def\ber#1{\begin{eqnarray}\label{#1} &&\nqq}%   left alignment
\def\eer{\end{eqnarray}}

\newcommand{\bear}[1]{\begin{eqnarray}\label{#1}}
\newcommand{\ear}{\end{eqnarray}}

\newcommand{\R}{ {\mathbb R} }

\newcommand{\fnm}{\footnotemark}
\newcommand{\fnt}{\footnotetext}

\begin{document}

\vspace{15pt}

 \begin{center}
 \large\bf

 On generalized Melvin solutions for  Lie algebras of rank $3$
  
 \vspace{15pt}

 \normalsize\bf
  S. V. Bolokhov\fnm[1]\fnt[1]{bol-rgs@yandex.ru}$^{, a}$
 and  V. D. Ivashchuk\fnm[2]\fnt[1]{ivashchuk@mail.ru}$^{, a, b}$

 \vspace{7pt}

 \it  (a) \  Institute of Gravitation and Cosmology,
 Peoples' Friendship University of Russia (RUDN University),
 6 Miklukho-Maklaya St.,  Moscow 117198, Russian Federation \\
 
 (b) \ \ \ Center for Gravitation and Fundamental
  Metrology,  VNIIMS, 46 Ozyornaya St., Moscow 119361, Russian Federation \\

 \end{center}
 \vspace{15pt}

 \small\noi

 \begin{abstract} 
Generalized Melvin  solutions for rank-$3$ Lie algebras $A_3$, $B_3$ and  $C_3$
 are considered. Any solution contains metric, three Abelian 2-forms and
 three scalar fields.  It is governed by three moduli functions $H_1(z),H_2(z),H_3(z)$
 ($z = \rho^2$ and $\rho$ is a radial variable), obeying
 three  differential equations with certain boundary conditions
 imposed. These functions are polynomials with powers $(n_1,n_2, n_3) = (3,4,3), (6,10,6), (5,8,9)$ for
 Lie algebras $A_3$, $B_3$, $C_3$, respectively. 
  The solutions depend upon integration constants $q_1, q_2, q_3 \neq 0$.
 The power-law asymptotic relations for  polynomials at large $z$ 
   are governed by  integer-valued $3 \times 3$ matrix $\nu$, which  coincides 
    with twice the inverse Cartan matrix $2 A^{-1}$ for Lie algebras  $B_3$ and  $C_3$, while
   in the $A_3$ case  $\nu = A^{-1} (I + P)$, where   $I$ is the identity matrix and
   $P$ is a  permutation matrix, corresponding to a generator of the $\mathbb{Z}_2$-group  
   of symmetry of the Dynkin diagram.   The duality identities for polynomials  and 
  asymptotic relations for solutions at large distances are obtained. 
  2-form flux integrals  over a  $2$-dimensional  disc of radius $R$ 
  and corresponding Wilson loop factors over a circle of radius $R$ are presented. 

\end{abstract}


\large 

 \section{Introduction}

  In this paper we deal with certain  generalizations of   Melvin solution \cite{Melv},
   which were presented  earlier in \cite{GI-09}. Originally,  model from \cite{GI-09}
  contains metric, $n$ Abelian 2-forms and  $l \geq n$ scalar fields.
    Here we consider a special solutions with $n = l =3$, governed by a $3 \times 3$ Cartan matrix  
  $(A_{i j})$ for simple Lie algebras of rank $3$: $A_3$, $B_3$ and  $C_3$.  
  The solutions from \cite{GI-09} are special case of the so-called generalized fluxbrane  
  solutions from \cite{Iflux}.  
  
  We note that Melvin's 
  original $4$-dimensional solution  describes the gravitational field of a magnetic flux tube. The
  multidimensional analog of such a flux tube, supported by a certain configuration of form fields, is 
  referred to as a fluxbrane (a ``thickened brane'' of magnetic flux). The appearance of fluxbrane 
  solutions was motivated by superstring/brane models and $M$-theory. A physical interest in such 
  solutions is that they may supply an appropriate background geometry for studying various processes 
  involving branes, instantons, Kaluza--Klein monopoles, pair production of magnetically charged 
  black holes and other configurations which can be studied via a special kind of Kaluza--Klein 
  reduction (``modding technique'') of a certain multidimensional model in the presence of $U(1)$ isometry. 
  It was shown, in particular, that Melvin's original solution ($F1$-fluxbrane) can be interpreted as a modding 
  of flat space in one dimension higher. This ``modding'' technique is widely used in construction of new
  solutions in supergravity models and also for various physical applications in superstring models and 
  M-theory. For generalizations of the  Melvin solution and  fluxbrane solutions see 
  \cite{Bron-79}-\cite{Ivas-Symmetry-17} and references therein.

  The generalized  fluxbrane solutions from  \cite{Iflux} are governed by 
  moduli functions   $H_s(z) > 0$ defined on the interval $(0, +\infty)$, where
   $z = \rho^2$ and $\rho$ is a radial variable. These functions 
  obey a set of  $n$ non-linear differential master equations governed by a matrix $(A_{s s'})$.
  (See equations (\ref{1.1}) below.)   These equations are equivalent to 
  Toda-like equations,  with  the following boundary conditions imposed:  $H_{s}(+ 0) = 1$,   $s = 1,...,n$.  

 It was assumed in \cite{GI-09} that $(A_{s s'})$ is a Cartan matrix for some 
 simple finite-dimensional Lie algebra $\cal G$ of rank $n$ ($A_{ss} = 2$ for all $s$).
 According to a conjecture  suggested in \cite{Iflux}, the
 solutions to master equations with the  boundary conditions imposed 
 are  polynomials: 
  \beq{1.3}
   H_{s}(z) = 1 + \sum_{k = 1}^{n_s} P_s^{(k)} z^k,
  \eeq
  where $P_s^{(k)}$ are constants. Here
 $P_s^{(n_s)} \neq 0$  and 
 \beq{1.4}
  n_s = 2 \sum_{s' =1}^{n} A^{s s'},
 \eeq 
 where we denote $(A^{s s'}) = (A_{s s'})^{-1}$.
 Integers $n_s$ are components  of the twice dual
 Weyl vector in the basis of simple (co-)roots \cite{FS}.
 
The set of fluxbrane polynomials $H_s$ defines a 
special solution to  open Toda chain equations \cite{K,OP} corresponding 
to  simple finite-dimensional Lie algebra $\cal G$ \cite{I-14}.
In refs. \cite{GI-09,GI} a program  (in Maple)  for calculation of these polynomials for
classical series of Lie  algebras  ($A$-, $B$-, $C$- and $D$-series) was suggested.
It was pointed out in \cite{Iflux} that the conjecture on polynomial structure of  $H_{s}(z)$ is valid for Lie  algebras of $A$- and $C$- series.

 One of the goals of this paper is to study interesting geometric properties of the solution considered. In
  particular, we prove the so-called duality property of fluxbrane polynomials for Lie algebras 
  of rank $3$,  which establishes a certain symmetry of the solutions
  under the inversion transformation $\rho \to 1/\rho$, which makes the model in tune with $T$-duality in 
  string models, and also can be mathematically understood in terms of the groups of symmetry 
   of Dynkin diagrams for the corresponding Lie algebras. In our case these groups of symmetry are 
   either identical ones (for Lie algebras $B_3$ and  $C_3$) or isomorphic to the group $\mathbb{Z}_2$ 
   (for $A_3$). These   duality identities may be used in deriving a $1/\rho$-expansion for solutions at large distances $\rho$.   The corresponding asymptotic behavior of the solutions is studied. 
   
   We note that analogous    analysis was performed recently for  the case of rank-2 Lie algebras: 
   $A_2$, $B_2 = C_2$, $G_2$  in \cite{BolIvas-R2-17}. In \cite{BolIvas-17}    the conjecture from  \cite{Iflux} was verified for the Lie algebra $E_6$ and certain duality relations for six $E_6$-polynomials were found.

Here we study  generalized Melvin solutions for  Lie algebras of rank $3$. The paper is organized as follows.
In Section 2 we present a generalized  Melvin solutions
from \cite{GI-09} for the case of three scalar fields and three 2-forms.
 In Section 3 we deal with the   solutions for the Lie algebras $A_3$,  $B_3$ and  $C_3$.
 We find duality relations for polynomials and present asymptotic relations for the 
 solutions. Here we also present  2-form flux integrals $\Phi^s(R) = \int_{D_{R}} F^s$ 
 and corresponding Wilson loop factors, where  $F^s$ are 2-forms and $D_{R}$ is  $2$-dimensional  
 disc of radius $R$. The flux integrals  have finite limits for $R = + \infty$ 
 \cite{Ivas-flux-17}.

\section{The Set Up and Generalized Melvin Solutions}

We consider a  model governed by the action
 \beq{2.1}
 S=\int d^Dx \sqrt{|g|} \biggl \{R[g]-
 \delta_{a b} g^{MN}\partial_M \varphi^{a} \partial_N \varphi^{b} - \frac{1}{2}
 \sum_{s =1}^{3}\exp[2 \vec{\lambda}_s \vec{\varphi}](F^s)^2 \biggr \},
 \eeq
 where $g=g_{MN}(x)dx^M\otimes dx^N$ is a metric,
 $\vec{\varphi} = (\varphi^a)\in \R^3$ is vector of scalar fields,
   $ F^s =    dA^s
          =  \frac{1}{2} F^s_{M N}  dx^{M} \wedge  dx^{N}$
 is a $2$-form,  $\vec{\lambda}_s = (\lambda_{s}^{a}) \in \R^3$ is dilatonic  coupling vector,
   $s = 1,2,3$; $a =1,2,3$.
 In (\ref{2.1})
 we denote $|g| =   |\det (g_{MN})|$, $(F^s)^2  =
 F^s_{M_1 M_{2}} F^s_{N_1 N_{2}}  g^{M_1 N_1} g^{M_{2} N_{2}}$.

 Here we deal with a family of exact
solutions to field equations corresponding to the action
(\ref{2.1}) and depending on one variable $\rho$. The solutions
are defined on the manifold
 \beq{2.2}
  M = (0, + \infty)  \times M_1 \times M_2,
 \eeq
 where $M_1 = S^1$ and
 $M_2$ is a $(D-2)$-dimensional Ricci-flat manifold. The solution
 reads \cite{GI-09}
 \bear{2.30}
  g= \Bigl(\prod_{s = 1}^{3} H_s^{2 h_s /(D-2)} \Bigr)
  \biggl\{  d\rho \otimes d \rho  +
  \Bigl(\prod_{s = 1}^{3} H_s^{-2 h_s} \Bigr) \rho^2 d\phi \otimes d\phi +
    g^2  \biggr\},
 \\  \label{2.31}
  \exp(\varphi^a)=
  \prod_{s = 1}^{3} H_s^{h_s  \lambda_{s}^a},
 \\  \label{2.32a}
  F^s =  q_s \left( \prod_{l = 1}^{3}  H_{l}^{- A_{s
  l}} \right) \rho d\rho \wedge d \phi,
  \ear
 $s, a = 1,2,3$, where  $g^1 = d\phi \otimes d\phi$ is a
  metric on $M_1 = S^1$ and $g^2$ is a  Ricci-flat metric of 
  signatute $(-,+, \dots, +)$ on
 $M_{2}$.  Here $q_s \neq 0$ are integration constants  
 ($q_s = - Q_s$ in notations of  \cite{GI-09}).

 The functions $H_s(z) > 0$, $z = \rho^2$, obey the master equations
\beq{1.1}
  \frac{d}{dz} \left( \frac{ z}{H_s} \frac{d}{dz} H_s \right) =
   P_s \prod_{l = 1}^{3}  H_{l}^{- A_{s l}},
  \eeq
 with  the following boundary conditions
 \beq{1.2}
   H_{s}(+ 0) = 1,
 \eeq
 where
 \beq{2.21}
  P_s =  \frac{1}{4} K_s q_s^2,
 \eeq
 $s = 1, 2,3$.  The boundary condition (\ref{1.2}) guarantees the absence 
 of a conic singularity (for the metric  (\ref{2.30})) for $\rho =  +0$.
 
 The parameters  $h_s$  satisfy the relations
  \beq{2.16}
  h_s = K_s^{-1}, \qquad  K_s = B_{s s} > 0,
  \eeq
 where
 \beq{2.17}
  B_{s l} \equiv
  1 +\frac{1}{2-D}+  \vec{\lambda}_{s} \vec{\lambda}_{l} ,
  \eeq
 $s, l = 1, 2, 3$.
 In the relations above we denote 
 \beq{2.18}
  (A_{s l}) = \left( 2 B_{s l}/B_{l l} \right).
 \eeq
 The latter is the so-called quasi-Cartan matrix. 
  
 It may be shown that if $(A_{s l})$ is a Cartan matrix 
 for a simple Lie algebra $\cal G$ of rank $3$
 there exists a set of vectors 
 $\vec{\lambda}_1,  \vec{\lambda}_2, \vec{\lambda}_3$  obeying (\ref{2.18}), 
 see Remark 1 in the next section.

The solution under consideration is as a special case of the
fluxbrane solution from \cite{Iflux,GIM}.

Thus, here  we deal with a multidimensional 
generalization of the  Melvin's solution \cite{Melv} for the case of three scalar fields and 
three $2$-forms.  The  Melvin's solution without scalar field corresponds to $D = 4$,
one $2$-form,  $M_1 = S^1$ ($0 < \phi <  2 \pi$),  $M_2 = \R^2$ and
$g^2 = -  dt \otimes dt + d x \otimes d x$.

\section{Solutions Related to Simple Lie Algebras of Rank $3$}

Here we deal with the solutions which corresponds to a simple  Lie algebras ${\cal G}$ of rank $3$, i.e. 
the matrix   $A =  (A_{sl})$ is coinciding with the Cartan matrix 
\beq{A.5}
     \left(A_{ss'}\right)=
     \begin{pmatrix}
      2 & -1 & 0 \\
      -1 & 2 & -1 \\
      0 & -1 & 2
     \end{pmatrix}\!,
  \quad
   \begin{pmatrix}
   2 & -1 & 0 \\
   -1 & 2 & -2 \\
   0 & -1 & 2
  \end{pmatrix}\!,
  \quad
  \begin{pmatrix}
     2 & -1 & 0 \\
     -1 & 2 & -1 \\
     0 & -2 & 2
    \end{pmatrix}
     \eeq
for ${\cal G} = A_3,  B_3,  C_3$, respectively. 

This matrix is described graphically by the Dynkin diagrams pictured on Fig. 1 for these three Lie algebras.
 \vspace{25pt}

 \setlength{\unitlength}{1mm}
 \begin{figure}[ph]
 \centering
 \begin{picture}(80, 0)
 \put(2,5){\circle*{2}}
 \put(12,5){\circle*{2}}
 \put(22,5){\circle*{2}}
 \put(1,5){\line(1,0){20}}
 \put(0,-0.5){$1$}
 \put(10,-0.5){$2$}
 \put(20,-0.5){$3$}
 
 \put(32,5){\circle*{2}}
 \put(42,5){\circle*{2}}
 \put(52,5){\circle*{2}}
 \put(32,5){\line(1,0){10}}
 \put(42,5.5){\line(1,0){10}}
 \put(42,4.5){\line(1,0){10}}
 \put(30,-0.5){$1$}
 \put(40,-0.5){$2$}
 \put(50,-0.5){$3$}
 \put(45,3.9){\large $>$}
 
 \put(62,5){\circle*{2}}
  \put(72,5){\circle*{2}}
  \put(82,5){\circle*{2}}
  \put(62,5){\line(1,0){10}}
  \put(72,5.5){\line(1,0){10}}
  \put(72,4.5){\line(1,0){10}}
  \put(60,-0.5){$1$}
  \put(70,-0.5){$2$}
  \put(80,-0.5){$3$}
  \put(75,3.9){\large $<$}
 \end{picture}
 \caption{The Dynkin diagrams for the Lie algebras $A_3$, $B_3$, $C_3$, respectively.\label{fig1}}
 \end{figure}
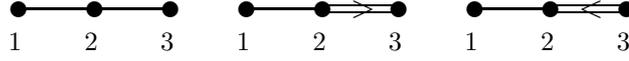
     
  Due to (\ref{2.16})--(\ref{2.18}) we get
    \begin{equation}
           \label{3.17}
    K_s =   \frac{D - 3}{D -2} +  \vec{\lambda}_{s}^2,
      \end{equation} 
    $h_s =  K_s^{-1}$,  and  
      \beq{3.18}
        \vec{\lambda}_{s} \vec{\lambda}_{l} = 
            \frac{1}{2} K_l A_{sl}  - \frac{D - 3}{D -2} \equiv G_{sl},
      \eeq    
     $s,l = 1, 2, 3$;  (\ref{3.17}) is a special case of  (\ref{3.18}). 
         
       It follows from  (\ref{2.16}), (\ref{2.18}) that 
       \beq{3.26}
       \frac{h_s}{h_l} = \frac{K_{l}}{K_{s}} = \frac{B_{ll}}{B_{ss}} = 
          \frac{B_{ls}}{B_{ss}} \frac{B_{ll}}{B_{sl}} = \frac{A_{ls}}{A_{sl}}
        \eeq
       for any $s \neq l$ obeying $A_{sl}, A_{ls} \neq 0$. This
       implies
       \beq{3.26K}
        K_1 =K_2 = K, \qquad K_3 = K, \ \frac{1}{2} K, \ 2K  
        \eeq
        or 
        \beq{3.26h}
        h_1 = h_2 = h, \qquad h_3 = h, \ 2h, \ \frac{1}{2} h ,   
        \eeq
        ($h = K^{-1}$) for ${\cal G} =  A_3,  B_3,  C_3$, respectively.

     \vskip 1em \noindent {\bf Remark 1.} 
     For large enough $K_1$ in (\ref{3.26}) there exist vectors 
     $\vec{\lambda}_s$  obeying   (\ref{3.18}) (and hence (\ref{3.17})).
     Indeed, the matrix $(G_{sl})$ is positive definite
     if $K_1 > K_{*}$,  where $K_{*}$ is some positive number. Hence there exists
     a matrix $\Lambda$, such that $\Lambda^{T} \Lambda = G$. We put
      $(\Lambda_{as}) = (\lambda_{s}^a)$ and get the set of vectors obeying
      (\ref{3.18}).
     
     \subsection{Polynomials}
     The set of moduli functions   $(H_1(z),H_2(z), H_3(z))$, obeying Eqs. (\ref{1.1}) and (\ref{1.2})
     with the matrix $A =  (A_{sl})$ from  (\ref{A.5}) 
     are polynomials with powers $(n_1,n_2,n_3) = (3,4,3), (6,10,6), (5,8,9)$ for
     ${\cal G} = A_3,  B_3,  C_3$, respectively.
     
     In what follows we list these polynomials. Here as in \cite{I-14} 
     we use  rescaled variables
     \beq{3.P}
        p_s = P_s/n_s.    
     \eeq
     %$s = 1,2,3$.
         
          {\bf $A_3$-case.} For the Lie algebra $A_3 \cong sl(4)$ 
          we have \cite{I-14,GI}
           \bear{A.6}
             &&H_1=1+3 p_1 z +3 p_1 p_2 z^2 +p_1 p_2 p_3 z^3,\\
            \label{A.7}
             &&H_2=1+4 p_2 z +(3 p_1 p_2+3 p_2 p_3)z^2 + 4 p_1 p_2 p_3 z^3 + p_1 p_2^2 p_3 z^4, \\
             &&H_3=1+3 p_3 z +3 p_2 p_3 z^2 + p_1 p_2 p_3 z^3.
            \ear

          {\bf $B_3$-case.}
           For the Lie algebra $B_3 \cong so(7)$  the fluxbrane polynomials read  \cite{GI}
\begin{eqnarray}\label{B.3}
&&\nq H_1=1+6 p_1 z +15 p_1 p_2 z^2 +20 p_1 p_2 p_3 z^3+15 p_1 p_2 p_3^2 z^4 +6 p_1 p_2^2 p_3^2 z^5+p_1^2 p_2^2 p_3^2 z^6,  \quad 
 \\
&&\nq H_2=1+10 p_2 z+ \left(15 p_1 p_2+30 p_2 p_3\right)z^2+\left(80 p_1 p_2 p_3+40 p_2 p_3^2\right)z^3 \nn
&&\;\; + \left(50 p_1 p_2^2 p_3+135 p_1 p_2 p_3^2+25 p_2^2 p_3^2\right)z^4 + 252 p_1 p_2^2 p_3^2 z^5\nn
&&\;\; +\left(25 p_1^2 p_2^2 p_3^2+135 p_1 p_2^3 p_3^2+50 p_1 p_2^2 p_3^3\right)z^6 +\left(40 p_1^2 p_2^3 p_3^2+80 p_1 p_2^3  p_3^3\right)z^7 \nn
&&\;\;+\left(30 p_1^2 p_2^3 p_3^3+15 p_1 p_2^3 p_3^4\right)z^8 +10  p_1^2 p_2^3 p_3^4z^9+ p_1^2 p_2^4 p_3^4 z^{10},
\\
&&\nq H_3=1+6 p_3z+15 p_2 p_3 z^2 +\left(10 p_1 p_2 p_3+10 p_2 p_3^2\right)z^3+15 p_1 p_2 p_3^2 z^4 \nn
&&\;\;+6 p_1 p_2^2 p_3^2 z^5 +p_1 p_2^2 p_3^3 z^6 . 
\end{eqnarray}
                         
           { \bf $C_3$-case. }
           For the Lie algebra $C_3 \cong sp(3)$ we obtain 
            (by using MATHEMATICA)  the following polynomials
           \vspace{10pt}
\begin{eqnarray}\label{C.3}
&&\nq H_1=1+5 p_1 z +10  p_1 p_2 z^2 +10 p_1 p_2 p_3 z^3 +5 p_1 p_2^2 p_3 z^4 + p_1^2 p_2^2 p_3 z^5 ,
\\
&&\nq H_2=1+8 p_2 z + \left(10 p_1 p_2+18 p_2 p_3\right)z^2+ \left(40 p_1 p_2 p_3+16 p_2^2 p_3\right) z^3 +70 p_1 p_2^2 p_3 z^4 \nn
&&\;\;+\left(16 p_1^2 p_2^2 p_3+40 p_1 p_2^3 p_3\right)z^5+\left(18 p_1^2 p_2^3 p_3+10 p_1 p_2^3 p_3^2\right)z^6 \nn
&&\;\;+8 p_1^2 p_2^3 p_3^2z^7 +p_1^2 p_2^4 p_3^2 z^8, 
\\
&&\nq H_3=1+9 p_3 z+36 p_2 p_3 z^2 +\left(20 p_1 p_2 p_3+64 p_2^2 p_3\right)z^3+\left(90 p_1 p_2^2 p_3+36 p_2^2 p_3^2\right)z^4 \nn
&&\;\; +\left(36 p_1^2 p_2^2 p_3+90 p_1 p_2^2 p_3^2\right)z^5 + \left(64 p_1^2 p_2^2 p_3^2+20 p_1 p_2^3 p_3^2\right)z^6 \nn
&&\;\;+ 36 p_1^2 p_2^3 p_3^2 z^7 +9 p_1^2 p_2^4 p_3^2 z^8 + p_1^2 p_2^4 p_3^3 z^9 . 
\end{eqnarray}
          
      Let us denote
      \begin{equation}
        \label{3e.5}
         H_s = H_s(z) = H_s(z, (p_i) ),
       \end{equation}
       where $(p_i) = (p_1,p_2,p_3)$.
      
      Due to  relations for  polynomials we have the following asymptotical behaviour 
      \begin{equation}
        \label{3e.6}
         H_s = H_s(z, (p_i) )  \sim \left( \prod_{l=1}^{3} (p_l)^{\nu^{sl}} \right) z^{n_s} \equiv 
         H_s^{as}(z, (p_i)),
       \end{equation}
       as $z \to \infty$. 
      
      Here $\nu = (\nu^{sl})$ is the integer valued matrix 
       \begin{equation}
         \label{3e.7}
         \nu =  \left(
         \begin{array}{cccccc}
          1 & 1 & 1 \\
          1 & 2 & 1 \\
          1 & 1 & 1
         \end{array}
         \right)\!,
         \quad  
         \left(
         \begin{array}{cccccc}
                         2 & 2 & 2  \\
                         2 & 4 & 4 \\
                         1 & 2 & 3  
                        \end{array}
                        \right)\!,
         \quad 
           \left(  
           \begin{array}{cccccc}
                         2 & 2 & 1  \\
                         2 & 4 & 2 \\
                         2 & 4 & 3 
                        \end{array}
                      \right)\!,
          \end{equation}
      for Lie algebras $A_3,  B_3,  C_3$, respectively.

     For Lie algebras  $B_3$ and $C_3$  we have 
      \begin{equation}
        \label{3e.8a}
      \nu = 2 A^{-1}, 
      \end{equation}
      where $A^{-1}$ is inverse Cartan matrix.    For the  $A_3$-case   
      the matrix $\nu$ is related to the inverse Cartan matrix as follows
      \begin{equation}
        \label{3e.8b}
        \nu = A^{-1} (I + P),
       \end{equation}
       where $I$ is $3 \times 3$ identity matrix and
      \begin{equation}
         \label{3e.9}
         P  = \left(
         \begin{array}{cccccc}
          0 & 0 & 1  \\
          0 & 1 & 0  \\
          1 & 0 & 0
          \end{array}
         \right).
         \end{equation}
      is permutation matrix. 
      This matrix corresponds to the permutation $\sigma \in S_3$ ($S_3$ is symmetric group)
      \begin{equation}
        \label{3.10}
         \sigma: (1,2,3) \mapsto (3,2,1),
       \end{equation}
       by the following relation  $P = (P^i_j) = (\delta^i_{\sigma(j)})$. 
       Here $\sigma$   is the generator of the group 
       $G = \{ \sigma, {\rm id} \}$ which is the group of symmetry of the Dynkin diagram (for $A_3$). 
       $G$ is isomorphic to the group $\mathbb{Z}_2$.
       
        We note that in all three cases we have 
        \begin{equation}
          \label{3e.10a}
           \sum_{l= 1}^3 \nu^{sl} = n_s,
         \end{equation}
         $s = 1, 2, 3$.
      
      \subsection{Duality relations}
       Let us denote $\hat{p}_i = p_{\sigma(i)} $ for the $A_3$ case
       and $\hat{p}_i = p_{i}$ for $B_3$ and $C_3$ cases, $i= 1,2,3$.         
       We call the ordered set $(\hat{p}_i)$ as \textit{dual} one to the ordered set $(p_i)$.
       By using the relations for  polynomials   we are 
       led  to the following identities which were verified by using MATHEMATICA. 
       
       \vskip 1em 
       \noindent {\bf Proposition.} {\em  The fluxbranes polynomials
        corresponding to Lie algebras $A_3$,
        $B_3$ and $C_3$ obey for all $p_i > 0$ and $z > 0$
        the identities
        \begin{equation}
          \label{3.12}
           H_{s}(z, (p_i) ) = H_s^{as}(z, (p_i)) H_s(z^{-1}, (\hat{p}_i^{-1})),
         \end{equation}
         $s = 1, 2, 3$. }
        
       We call  relations (\ref{3.12})  as duality ones.
               
       \subsection{Fluxes}
       
       Now let us consider the oriented $2$-dimensional manifold 
      $M_{R} =(0, R)  \times S^1$, $R > 0$. The flux integrals
       \beq{3.19}
        \Phi^s(R) = \int_{M_{R}} F^s =
         2 \pi \int_{0}^{R} d \rho \rho {\cal B}^s ,
       \eeq    
       where 
        \begin{equation}
           \label{3e.16}
           {\cal B}^s =   q_s  \prod_{l = 1}^{3}  H_{l}^{- A_{s l}}.
        \end{equation} 
       
       Total flux integrals $\Phi^s = \Phi^s(+ \infty)$  are convergent.        
        Indeed, due to polynomial assumption  (\ref{1.3}) we have 
      \beq{3e.20}
       H_s \sim C_s \rho^{2n_s},   \qquad C_s = \prod_{l = 1}^{3} (p_l)^{\nu^{sl}},
            \eeq 
      as $\rho \to + \infty$. From (\ref{3e.16}), (\ref{3e.20}) and 
      the equality $\sum_{1}^{n} A_{s l} n_l = 2$,
      following from (\ref{1.4}), we get
      \beq{3.21}
      {\cal B}^s \sim  q_s C^s \rho^{-4},  \quad  
       C^s = \prod_{l = 1}^{3}  p_{l}^{- (A \nu) _{s}^{ \ l}},
      \eeq 
       and hence the integral (\ref{3.19}) is convergent for any $s =1,2,3$.   
       
        Due to (\ref{3e.8b})  and (\ref{3.21}) we get for the $A_3$-case 
          \beq{3e.23}
              C^s =  \prod_{l = 1}^{3} p_l^{-(I+P)_{s}^{\ l}} = 
             \prod_{l = 1}^{3} p_l^{- \delta^l_s - \delta^l_{\sigma(s)} } = 
                   p_s^{-1} p_{\sigma(s)}^{-1}.               
          \eeq 
            Using (\ref{3e.8a}) and  (\ref{3.21}) we obtain for Lie algebras $B_3$ and $C_3$            
          \beq{3e.23cg}
           C^s =   p_s^{-2}.
          \eeq
                             
       Now we calculate  $\Phi^s(R)$.
        By using  master equations (\ref{1.1})    we find
      \bear{3.24}
      \int_{0}^{R} d \rho \rho {\cal B}^s =  q_s P_s^{-1}
      \frac12 \int_{0}^{R^2} d z  
      \frac{d}{dz} \left( \frac{ z}{H_s} \frac{d}{dz} H_s \right)
      \\ \nonumber
      =     \frac12 q_s P_s^{-1}               
                \frac{R^2 H_s^{'}(R^2) }{H_s (R^2)},          
      \ear
   where $H_s^{'} = dH_s/dz$.
   
    Using  (\ref{3.19}) we obtain 
    \beq{3.24f}
         \Phi^s(R)  =   4 \pi  q_s^{-1} h_s                
                    \frac{R^2 H_s^{'}(R^2) }{H_s (R^2)}.          
    \eeq

     The manifold $M_{*} =(0, + \infty)  \times S^1$ is isomorphic to
     the manifold $\R^2_{*} = \R^2 \setminus \{ 0 \}$.  The solution under consideration may be understood 
     as defined on the manifold $\R^2_{*} \times M_2$,  where coordinates $\rho$, $\phi$ are polar coordinates in a domain of  $\R^2_{*}$: $x = \rho \cos \phi$ and
     $y = \rho \sin \phi $, where $x, y$ are standard coordinates of $\R^2$. It was shown  
     in \cite{Ivas-flux-17} that  there exist forms $A^s$ obeying $F^s = dA^s$ 
     which are globally defined on $\R^2$.
                            
     Let us consider a 2D oriented manifold (disk) $D_R = \{ (x,y): x^2 + y^2 \leq R^2 \}$ with the boundary
     $\partial D_R = C_R = \{ (x,y): x^2 + y^2 = R^2 \}$. $C_R$ is a circle of radius $R$. It is
     an 1D oriented manifold  with the orientaion (inherited from that of $D_R$) obeying the relation $\int_{C_R} d\phi = 2 \pi$.  
    Using the Stokes theorem we get
      \beq{3.24fA}
                 \Phi^s(R)  =   \int_{M_{R}} F^s = \int_{D_{R}} d A^s = \int_{C_{R}}  A^s.          
      \eeq
      
            Using the definition of Abelian Wilson loop (factor) we get 
       \beq{3.W}
              W^s(C_R)  =  \exp (i \int_{C_{R}}  A^s ) = \exp (i  \Phi^s(R) ).          
       \eeq

  Relation (\ref{1.3}), (\ref{3.24f})  implies (see (\ref{2.21}))
      \beq{3.25}
       \Phi^s =  \Phi^s (+ \infty) =   4 \pi n_s q_s^{-1} h_s,  
      \eeq
   $s =1, 2,3$.  Any (total) flux $\Phi^s$ depends upon one integration constant  $q_s \neq 0$, 
  while the integrand form $F^s$ depends upon all constants: $q_1, q_2, q_3$.
  As a consequence, we obtain finite limits
  \beq{3.WL}
    \lim_{R \to + \infty} W^s(C_R)  =  \exp (i \Phi^s).          
   \eeq

     We get in the $A_3$-case
      \beq{3.A2f}
     (q_1 \Phi^1, q_2 \Phi^2, q_3 \Phi^3) =   4 \pi h  (3,4,3),
      \eeq
      where $h_1 = h_2 = h_3 = h$. 
             
      In the $B_3$=case we find 
      \beq{3.Cf}
      (q_1 \Phi^1, q_2 \Phi^2, q_3 \Phi^3) =   4 \pi  ( 6 h_1, 10 h_2, 6 h_3) =
      4 \pi h ( 6, 10, 12),
      \eeq
     where $h_1 =  h_2 = h$, $h_3 = 2h $.
  
  For $C_3$-case we are led to relations 
   \beq{3.Gf}
   (q_1 \Phi^1, q_2 \Phi^2, q_3 \Phi^3) =   4 \pi  ( 5 h_1, 8 h_2, 9 h_3) =
        4 \pi h ( 5, 8, 9/2),
    \eeq
   where $h_1 =  h_2 = h$, $h_3 = \frac{1}{2} h $.  (In all examples relations (\ref{3.26h})
   are used.)
  
    We note that  for $D =4$ and 
   $g^2 = -  dt \otimes dt + d x \otimes d x$, $q_s$ is coinciding  with 
  the value of the $x$-component of the $s$-th magnetic field on the axis of symmetry, $s =1,2,3$. 
     
   \subsection{Asymptotic relations} 
     The asymptotic relations for the solution under consideration for 
      $\rho \to + \infty $ read
           \bear{3e.26}
            g_{as} = \Bigl(\prod_{l = 1}^{3} p_l^{a_l} \Bigr)^{2/(D-2)} \rho^{2A}
            \biggl\{ d\rho \otimes d \rho  \qquad \\ \nonumber
             +
            \Bigl(\prod_{l = 1}^{3} p_l^{a_l } \Bigr)^{-2} 
            \rho^{2 - 2A (D-2)} d\phi \otimes d\phi +  g^2 \biggr\},
            \\  \label{3e.27}
            \varphi^a_{as}=  \sum_{s = 1}^{3} h_s \lambda_{s}^a 
            (\sum_{l = 1}^{3} \nu^{sl} \ln p_l + 2 n_s \ln \rho ),
            \\  \label{3e.28}
            F^s_{as} = q_s  p_s^{-1} p_{\theta(s)}^{-1} \rho^{-3}  d\rho \wedge d \phi,
            \ear       
            $a, s =1,2,3$,
            where 
             \beq{3.25}
              a_l = \sum_{s =1}^3 h_s \nu^{sl}, \qquad   A =  2 (D-2)^{-1} \sum_{s = 1}^{3} n_{s} h_s, 
             \eeq
         and  in (\ref{3e.28}) we put $\theta = \sigma $ for ${\cal G} = A_3$, 
         and $\theta = id $ for ${\cal G} = B_3,  C_3$.
         In derivation of asymptotic relations Eqs. (\ref{3e.10a})  (\ref{3e.20}), (\ref{3e.23}) 
         were used.   We note that   for ${\cal G} = B_3,  C_3$ the asymptotic value of form
          $F^s_{as}$ depends upon $q_s$, $s = 1,2,3$, while in the $A_3$-case  $F^s_{as}$ 
          depends upon  $q_1$ and $q_3$ for $s = 1,3$, and $F^2_{as}$ depends upon  $q_2$.
                       
     We  note also that  by putting $q_1 = 0$ we get the Melvin-type solutions corresponding 
     to Lie algebras $A_2$ and $B_2 = C_2$, which were analyzed in \cite{BolIvas-R2-17}.
     (The case of the Lie algebra $G_2$ was also considered there.)

    \vskip 1em \noindent {\bf Remark 2.} Relations (constraints) on dilatonic coupling vectors  
    (\ref{2.17}), (\ref{2.18}) appear also for dilatonic black hole (DBH) solutions which   
     are defined on the manifold
    \beq{2b.2}
      M = (R_0, + \infty) \times (M_0 = S^2) \times (M_1 = \R) \times M_2,
     \eeq
     where $R_0 = 2 \mu > 0$ and $M_2$ is a  Ricci-flat manifold. These DBH solutions on $M$ from 
     (\ref{2b.2})
     for the model under consideration may be extracted from general black brane solutions, see  
     refs. \cite{Ivas-Symmetry-17,I-14,IMp3}. They read: 
     \bear{2b.30}
      g= \Bigl(\prod_{s = 1}^{3} {\bf H}_s^{2 h_s /(D-2)} \Bigr)
      \biggl\{  f^{-1} dR \otimes d R  + R^2 g^0  -
      \Bigl(\prod_{s = 1}^{3} {\bf H}^{-2 h_s} \Bigr) f dt \otimes dt +
        g^2  \biggr\},
     \\  \label{2b.31}
      \exp(\varphi^a)=
      \prod_{s = 1}^{3} {\bf H}^{h_s  \lambda_{s}^a},
     \\  \label{2b.32a}
      F^s = - Q_s R^{-2} \left( \prod_{l = 1}^{3}  {\bf H}_{l}^{- A_{s
      l}} \right)  dR \wedge d t,
      \ear
     $s,a = 1,2,3$, where $f = 1 - 2\mu R^{-1}$, $g^0$ is the standard 
      metric on $M_0 = S^2$ and $g^2$ is a  Ricci-flat metric of 
      signatute $(+, \dots, +)$ on $M_{2}$.  Here $Q_s \neq 0$ are integration constants  
     (charges).
         The functions ${\bf H}_s = {\bf H}_s (R) > 0$ obey the master equations
    \beq{1b.1}
      R^2 \frac{d}{dR} \left( f \frac{ R^2}{{\bf H}_s} \frac{d}{dR} {\bf H}_s \right) =
       B_s \prod_{l = 1}^{3}  {\bf H}_{l}^{- A_{s l}},
      \eeq
     with  the following boundary conditions on the horizon and at infinity imposed:
     \beq{1b.2}
       {\bf H}_{s}(R_0 + 0) =  {\bf H}_{s0} > 0, \qquad {\bf H}_{s}(+ \infty) = 1,
     \eeq
     where
     \beq{2b.21}
      B_s =  - K_s Q_s^2,
     \eeq
     $s = 1, 2,3$. Here relations (\ref{2.16}) are also valid. For Lie algebras 
     of rank 3 the functions ${\bf H}_s$ are polynomials with respect to $R^{-1}$, which may be 
     obtained (at least for small enough $q_s$) from fluxbrane polynomials $H_s(z)$ 
     presented in this paper. See  \cite{I-14}.

  \section{Conclusions}
  
    Here we have considered  generalizations of the Melvin's
    solution corresponding to  simple finite-dimensional Lie algebras of rank $3$:
    ${\cal G} = A_3,  B_3,  C_3$. 
    Any solution is governed by a set of $3$  fluxbrane polynomials $H_s(z)$, $s =1,2,3$. 
    These polynomials  define special solutions to open Toda chain equations corresponding 
    to the Lie algebra  ${\cal G}$.
     
    The polynomials $H_s(z)$ depend also upon parameters $q_s$, which   
    are coinciding for $D =4$  (up to a sign) with the values  of  colored 
    magnetic fields on the axis of symmetry. 
     
    The  duality identities for polynomials are found.  
    These  identities may be used  
   in deriving  $1/\rho$-expansion for solutions at large distances $\rho$, e.g. 
   for asymptotic relations  which are presented   in the paper.  
             
   We have  presented 2D flux integrals $\Phi^s(R) = \int_{D_R} F^s$ ($s =1, 2,3$) on a disc $D_R$
   of radius $R$ and a corresponding Wilson loop factors $W^s(C_R)$ over a circle $C_R$ of radius $R$.
   Any total flux  $\Phi^s (\infty)$  depends only upon one parameter $q_s$, 
   while the integrand  $F^s$ depends upon all parameters $q_1, q_2, q_3$.
           
  Another possible application of the solutions considered here is to study
   cosmological analogs of such solutions with phantom scalar fields. Such
   cosmological solutions for  Lie algebras of rank  $3$ were analysed in  \cite{Gol-10}
    for a special choice of the parameters $q_s$.

\section*{Acknowledgments}

 The publication has been  prepared with the support of the ``RUDN University Program 5-100''.
 It was also partially supported by the  Russian Foundation for Basic Research,  grant  Nr. 16-02-00602.

\end{document}